\documentclass[%
 aip,
 amsmath,amssymb,
 reprint,%
]{revtex4-1}

\usepackage[english]{babel}
\usepackage[normalem]{ulem}
\usepackage{graphicx}% Include figure files
\usepackage{dcolumn}% Align table columns on decimal point
\usepackage{bm}% bold math
\usepackage{xcolor}
\usepackage{tensor}
\usepackage{comment}
\newcommand{\eps}{\varepsilon}
\DeclareMathOperator{\Tr}{Tr}
\begin{document}

%\preprint{APS/123-QED}

\title{Methods for Creation and Linear Elastic Response Analysis of Packings of Semi-flexible Soft Polymer Chains and Molecules}% Force line breaks with \\
%\thanks{A footnote to the article title}

\author{R. C. Dennis}
\affiliation{Department of Physics and Astronomy, University of Pennsylvania, 209 South 33rd Street, Philadelphia, Pennsylvania 19104, USA}%\noaffiliation

\date{\today}% It is always \today, today,

\begin{abstract}
From understanding the sand on the beach to the foam on your beer, soft sphere simulations have been crucial to the study of the amorphous world around us. However, many of the materials we interact with on a daily basis aren't comprised of individual grains, but complex molecules and chains of polymers. By extending the soft sphere model to a model of linked spheres, we can learn more about the materials we interact with on a daily basis. In this methods paper, I show how one can find and study the physical properties of packings of flexible chains, rigid molecules, and everything in between. In addition to describing the energy landscape of these materials, these methods describe how to shear stabilize polymer packings, classify them in the jamming hierarchy, describe their elastic properties, and much more. This simple modification to soft sphere simulations has the potential to yield new discoveries surrounding glasses.
\end{abstract}

\maketitle
\begin{comment}
Outline:
1. Introduction: Talk about how this is more useful than sphere packings. Applications in biology and chemistry for small polymers with rigid and flexible degrees of freedom. Constraining the degrees of freedom
2. Figures:
x	a. Demonstrating the definition of a link
x	b. A minimized packing in 2 and 3D
x	c. Entertwined chains concern
	d. Floppy chain identification in 2D
	e. Shear and bulk moduli for randomly linked shear 					stabilized monomer packings as fct of link percent
3. Generating the Packings
4. Testing for strict jamming
5. Computing the moduli

\end{comment}
% \ec{Start text}
% 
% %\tableofcontents
\section{Introduction}
Glasses, grains, foams, sand, particulates, and colloidal suspensions are all examples of amorphous systems that undergo complicated phase transitions. These transitions have been studied extensively through simulations with packings of soft, athermal, frictionless spheres that interact through a one-sided contact potential. The success of this simplified model in furthering our understanding of glasses and jamming cannot be overstated~\cite{liu_jamming_1998,liu_jamming_2010,dagois-bohy_soft-sphere_2012,charbonneau_jamming_2015, charbonneau_universal_2016,lin_evidence_2016, morse_echoes_2017, dennis_jamming_2020}. However, many glassy systems are not comprised of soft spheres, but molecules and polymer chains that interact in more complex ways~\cite{ichihara_thermodynamic_1971, cangialosi_dynamics_2014, monnier_reaching_2021, wu_new_2021}.

While the use of soft sphere packings will not and should not subside in the future, we can further our understanding of glassy systems if we begin simulating packings of soft chains of spheres. I show how this can be done through a simple modification to code for soft sphere simulations.

The importance of polymer studies ranges from the biology of complex biomolecules such as proteins and chains of DNA~\cite{watson_molecular_1953, boyer_overview_2011, numata_how_2020} to the polyethylene that is omnipresent in our daily lives~\cite{royer_production_2018, ghatge_biodegradation_2020}. Packings of chains and polymers have been explored in experimental systems as well as in simulations~\cite{zou_packing_2009, barrat_molecular_2010,karayiannis_spontaneous_2012, hoy_jamming_2017, gartner_modeling_2019, mavrantzas_using_2021}.

Here a novel method is proposed that can be implemented and adapted to arbitrary cluster types and constraints. While real polymers are comprised of components joined by various types of interparticle forces, the work in this manuscript is done for the limiting case where the interparticle linkages are much stronger than the interaction potential between unlinked particles. The methods in this manuscript are focused specifically on soft chains of spheres and molecules which interact via a one-sided contact potential. However, the methods can be generalized to include more sophisticated interactions. Beyond the simple simulations of clusters of rigid molecules and flexible polymeric chains, one can also simulate clumping, aggregation, and the cementing of particles. 

In this manuscript, I demonstrate how to create overjammed and critically jammed packings of arbitrarily defined semi-rigid clusters of spheres. It is further explained how these can be prepared and shear stabilized. I also describe how to find features of the packing such as the normal modes, rattling clusters, classification in the jamming hierarchy, and elastic moduli. This simple extension to the already successful soft sphere packing model opens the door for many avenues of exploration in the physical properties of glassy systems.

\begin{figure}[]
\includegraphics[width=0.475\textwidth]{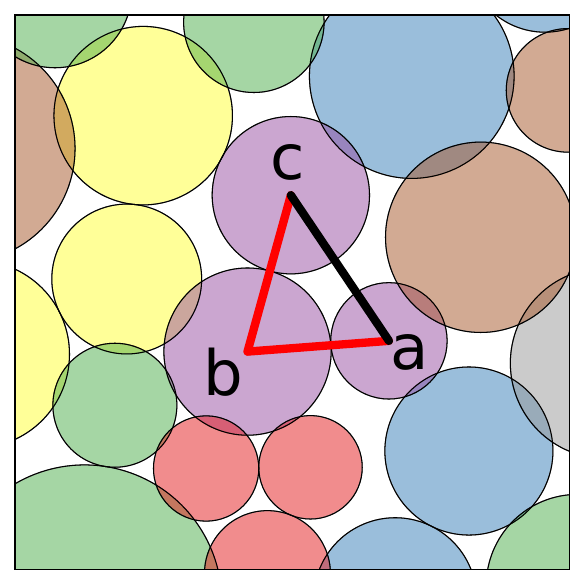}
\caption{An example of a set of links in a packing. The red lines and the black line are all links, chosen to keep the purple cluster rigid. The red links join the particles together to form a cluster of three particles while the black link fixes the bond angle preventing the cluster from deforming. This rigid cluster only has three degrees of freedom: two translational and one rotational. However, three independent, unconstrained particles have six degrees of freedom. Adding the three links above effectively removes three degrees of freedom.}\label{fig:links}
\end{figure}

\section{Generating the Packings}
Below a procedure is presented for generating packings of soft polymer chains and molecules that are force balanced in a local energy minimum through the use of constraints.
\subsection{Generating Polymers: Links}
The polymer packings are comprised of individual sets (or clusters) of spheres. The individual spheres in each cluster interact via a potential, $U,$ that is a function of their normalized overlaps, $\xi.$ To define these clusters, consider linkages between particles $i$ and $j$ of a fixed length. These are referred to as links.

By defining links, one is able to not only constrain the particles to be connected in arbitrary ways, but one can also fix the bond angles between particles. For example, given a triplet of circles, $a,~b,$ and $c$ connected via links $ab$ and $bc,$ the bond angle can be fixed by creating a link of some fixed length $l$ between $a$ and $c$ (see Figure~\ref{fig:links}).

\subsection{The Orthonormalized Constraint Basis}
Applying an arbitrary force to each of the particles in a packing would break their links and violate the constraints. The approach taken to prevent this is to project out the part of the force vector that violates the constraints to first order. Performing this projection will result in the forces that the chains actually experience and cause the appropriate cluster motions and torques in the simulation. Finding the projection requires finding an orthonormalized constraint basis.

In order to derive the orthonormal constraint basis, first consider the vector between two particles $i$ and $j$ that share a link, $\vec{x}_{ij}.$ Because we are working in periodic boundary conditions for arbitrary lattice vectors, I define the lattice vector matrix, $\Lambda,$ to be a $d\times d$ matrix where the columns are the lattice vectors. By considering a vector of integer lattice coordinates $\vec{z}_{ij} \in \mathbb{Z}^d,$ 
\begin{align}
x_{ij}^{\alpha}=x_j^\alpha-x_i^\alpha + \tensor{\Lambda}{^\alpha_\gamma}z_{ij}^{\gamma}
\end{align}
where $\alpha$ and $\gamma$ index the coordinates. It is also required that $\vec{z}_{ij}$ be an integer lattice vector which minimizes the norm of $x.$ This is the minimum image convention. For particles in contact, this means that $\vec{z}_{ij}$ is any integer vector that causes particle images $i$ and $j$ to be in contact. In general, this is degenerate.
If there is a link between particles $i$ and $j,$ our constraint is
\begin{align}
c_{\langle ij\rangle}=\sqrt{\sum_{\alpha}\left(x_{ij}^\alpha\right)^2}-l_{\langle ij\rangle}=0
\end{align} where $l_{\langle ij \rangle}$ is the constant length of link $\langle ij\rangle.$ 
Defining 
\begin{align}
\hat{\tau}_{ij}^{\alpha}\equiv\frac{x_{ij}^\alpha}{\sum_{\beta}\left(x_{ij}^\beta\right)^2}
\end{align}
to be the normalized distance vector between particles $i$ and $j,$ the constraint Jacobian is
\begin{align}
\Omega_{k\alpha\langle ij\rangle}=\hat{\tau}_{ij}^{\alpha}\left(\delta_{jk}-\delta_{ik}\right)
\end{align}
such that the columns of this matrix are given by the derivatives of $c.$
If the packing does not contain degenerate links, then the constraint Jacobian has the property that the number of columns is equal to the rank. In practice, it is very easy to remove degenerate links, so to simplify computations and derivations, we assume there are no degenerate links.

This constraint Jacobian becomes very powerful in determining the forces on the unconstrained degrees of freedom. Because links in the same cluster share particles, the columns of this matrix will in general not be orthonormal. One can perform QR decomposition~\cite{trefethen_numerical_1997} on the constraint Jacobian to find an orthonormal basis for the constraints. QR decomposition can be fairly computationally intensive, so a parallelized modified Gram-Schmitt process is employed~\cite{kerr_qr_2009} on each of the clusters individually. Since no two independent clusters share degrees of freedom, their columns will be automatically orthogonal. After performing the decomposition on our constraint Jacobian, the orthonormalized constraint basis, $\Omega_\textrm{o}$ is found. This matrix can be used to constrain the forces applied to the packing. Given an arbitrary force vector, $\vec{F},$ which acts on the particle positions, one can project out the part of the vector that lies along the constraints. This projection gives the constrained force vector:
\begin{align}
\vec{F}_{\textrm{cons}}&=\vec{F}-\Omega_\textrm{o}\Omega_\textrm{o}^T\vec{F}\nonumber\\
&=\left(\mathbf{1}-\Omega_\textrm{o}\Omega_\textrm{o}^T\right)\vec{F}.\label{constrainedForce}
\end{align}

For a system of rigid clusters, this constrained force removes the need to consider torques as it performs the correct angular rotation to first order. Similarly, for clusters with free bond angles, this alters the bond angles appropriately. It is worth noting that equation~\ref{constrainedForce} does not apply only to forces, but other variables as well. For example, one can find a constrained velocity in the same manner.
\subsection{The Lagrangian}
The constrained displacement of the particles is the vector that is closest to the proposed unconstrained displacement and does not violate the constraints, $c_{\langle ij\rangle}.$ This can be shown by considering the Lagrangian. Let the vector of constraint functions be given by $\vec{c}(\vec{x})$ for configuration $\vec{x}.$ Further, let the unconstrained displacement vector be given by $\Delta \vec{x}$ and the constrained displacement vector be given by $\Delta \vec{y}.$ Practically, this can be achieved by simply flattening the matrix of positions so that $\Delta\vec{x}$ and $\Delta \vec{y}$ both have length $Nd.$ We wish to minimize the distance from $\Delta \vec{x}$ and $\Delta \vec{y},$ so the corresponding Lagrangian is
\begin{align}
\mathcal{L}\left(\Delta\vec{y}, \vec{\lambda}\right)&=\left(\Delta \vec{y}-\Delta \vec{x}\right)^2+\vec{\lambda}^T\vec{c}(\vec{x}+\Delta\vec{y})\nonumber\\
&\approx \left(\Delta \vec{y}-\Delta \vec{x}\right)^2+\vec{\lambda}^T\Omega^T\Delta\vec{y}
\end{align}
where $c$ is expanded to first order and $c(\vec{x})=0.$
We wish to find the minimum of this Lagrangian, so we require
\begin{align}
0 &\stackrel{!}{=} \frac{\partial \mathcal{L}}{\partial \Delta \vec{y}^T} \nonumber \\
&\stackrel{!}{\approx}2\left(\Delta \vec{y}-\Delta \vec{x}\right)+\Omega \vec{\lambda}.\label{eqn:derivLag}
\end{align}
If we define the QR decomposition of $\Omega$ as $\Omega=\Omega_0 R_0,$ then we can multiply equation~\ref{eqn:derivLag} by $\Omega_0\Omega_0^T$ to find that
\begin{align}
0&=-2\Omega_0\Omega_0^T\Delta\vec{x}+\Omega\vec{\lambda}
\end{align}
where we used the fact that $\Omega_0^T\Delta\vec{y}=0.$
Combining this with equation~\ref{eqn:derivLag} gives the relationship,
\begin{align}
\Delta\vec{y}=\left(\mathbf{1}-\Omega_0\Omega_0^T\right)\Delta \vec{x}.
\end{align}  
For our purposes, the displacement of elements in our system must have a linear relationship with both the forces and velocities, so we can use this projection to ensure our displacements maintain the constraints to first order at any given step.
\subsection{Higher Order Corrections}\label{sec:hoc}
The previously described constraint method is only correct for infinitesimal perturbations. As the goal is to perform a quench on these clusters, the non-linear contributions will accumulate over the course of the simulation causing our constraints to be violated. To combat this, the constraints are periodically reaffixed. This can be achieved numerically by employing the Newton-Rhapson method~\cite{verbeke_newtonraphson_1995}. Given our initial positions are $\vec{x}_0,$ the following recursive equation can be solved iteratively:
\begin{align}
\Omega^T(\vec{x}_{k})\left[\vec{x}_{k+1}-\vec{x}_{k}\right]=-\vec{c}(\vec{x}_{k})
\end{align}
for $\vec{x}_{k+1}$ where $k$ represents the iteration number. This equation can be solved using a method such as Gaussian elimination~\cite{trefethen_numerical_1997}. This algorithm is continued until iteration $n$ where $\left|\vec{c}(\vec{x}_n)\right|<p$ for the desired precision $p.$ When starting in a position that is very close to satisfying the constraints, this algorithm should terminate after just a few steps.

This now fully defines a method for generating semi-flexible soft polymer packings, however if one also wants to probe the linear elastic response, they should do so for packings that are stable with respect to strains.

\begin{figure*}[]
\includegraphics[width=1\textwidth]{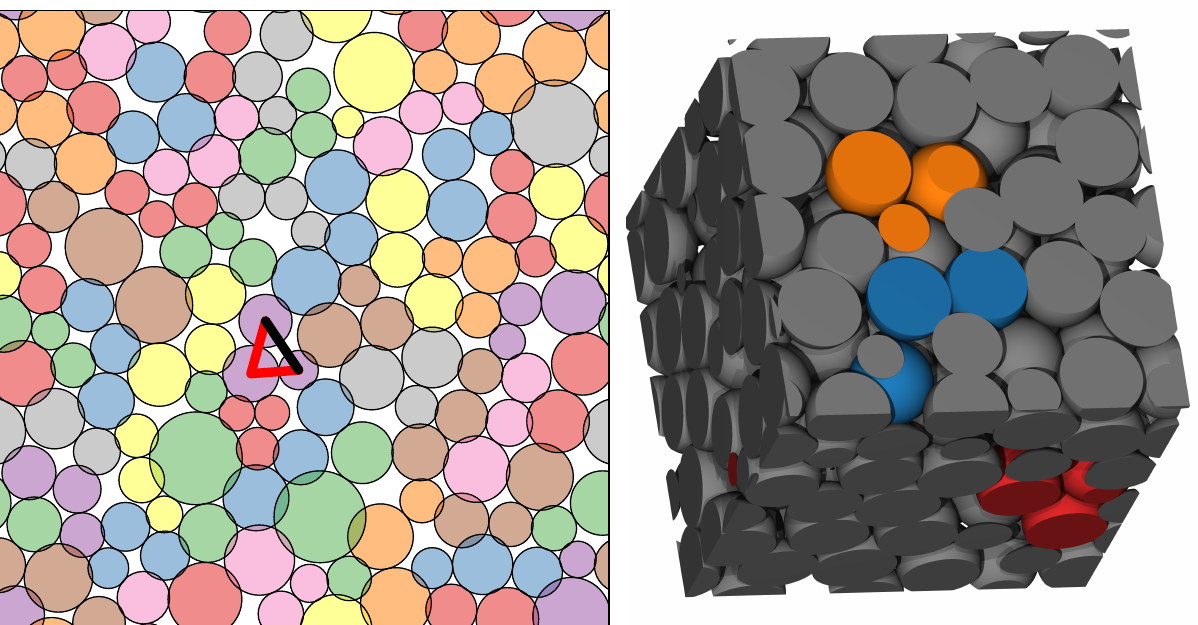}
\caption{Polymer packings made up of $50$ clusters of three particles with fixed bond angles in 2D (left) and 3D (right). For visualization purposes, only three of the rigid clusters are shown in the three dimensional packing.}\label{fig:minimized}
\end{figure*}

\subsection{Shear Stabilization}
While true stress-strain isotropy is typically reserved only for certain perfect crystalline packings, the amorphous nature of large thermal systems of grains and polymers causes them to be approximately isotropic as well. However, this is only true if they are shear stabilized. That is, if the packing is not at a minimum with respect to all strain degrees of freedom, then there exists a strain which when applied causes the energy of the packing to decrease. This violates the isotropic assumption which means that the packings cannot be described with elastic moduli. While plenty of excellent research has been done on elastic response in systems which are not shear stabilized~\cite{liao_stress-strain_1997, manning_vibrational_2011-1, manning_vibrational_2011, banigan_chaotic_2013, rainone_pinching_2020}, the desire for shear stabilization is often warranted nonetheless~\cite{donev_jamming_2004, dagois-bohy_soft-sphere_2012}.

Shear stabilization is achieved by minimizing the energy of a packing with respect to both position and strain degrees of freedom. In other words, we also consider the packing's lattice vectors to be subject to change. Given the position of a node, $\vec{p}_0,$ one can describe its lattice images as 
\begin{align}
\vec{p}=\vec{p}_0+\Lambda\vec{z}.\label{eq:pz}
\end{align}
A perturbation of the lattice vectors, $\Delta \Lambda,$ results in
\begin{align}
\Delta \vec{p}=\left(\Delta\Lambda\right)\vec{z}.
\end{align}
Solving equation~\ref{eq:pz} for $\vec{z}$ and making a substitution,
\begin{align}
\Delta\vec{p}=\left(\Delta\Lambda\right)\Lambda^{-1}\left(\vec{p}-\vec{p}_0\right).
\end{align}
Therefore, from the definition of the strain matrix,
\begin{align}
\mathbf{\eps}\equiv\nabla_{\vec{p}}\left(\Delta \vec{p}\right)=\left(\Delta\Lambda\right)\Lambda^{-1}.
\end{align}
It's also important to notice that the strain matrix must be symmetric; this can be achieved by defining it's derivatives as
\begin{align}
\frac{\partial \mathbf{\eps}_{ij}}{\partial \mathbf{\eps}_{kl}}\equiv \frac{1}{2}\left(\delta_{ik}\delta_{jl}+\delta_{il}\delta_{jk}\right).\label{eq:symmProp}
\end{align}
It is crucial that the positions of particles be a function of the affine strain, so one can then find the distance between particle $i$ and $j$ as a function of position and strain,
\begin{align}
\psi_{ij}^{\alpha}=\left(\tensor{\delta}{^\alpha_\beta}+\tensor{\eps}{^\alpha_\beta}\right)\left(x_j^{\beta}-x_i^{\beta}+\tensor{\Lambda}{^\beta_\gamma}z_{ij}^{\gamma}\right).
\end{align}

The constraints, $c_{\langle ij \rangle},$ and constraint Jacobian, $\Omega,$ can also be redefined as functions of the strain degrees of freedom. This means that
\begin{align}
c_{\langle ij\rangle}(\eps)=\sqrt{\sum_{\alpha}\left(\psi_{ij}^{\alpha}(\eps)\right)^2}-l_{\langle ij \rangle}.
\end{align}
The columns of $\Omega$ are comprised of derivatives of the constraints with respect to all of the degrees of freedom, positional and strain. The derivatives are evaluated at the current positions of the particles and the current strain of the packing. Because the positions and lattice vectors are updated at every step of the minimization protocol, the constraint derivates can be evaluated at zero strain. One can derive that
\begin{align}
\frac{\partial c_{\langle ij\rangle}}{\partial x_{k}^{\alpha}}\bigg|_{\eps=0}=\hat{\tau}_{ij}^{\alpha}\left(\delta_{jk}-\delta_{ik}\right)
\end{align}
as before. However,
\begin{align}
\frac{\partial c_{\langle ij\rangle}}{\partial \tensor{\eps}{^a_b}}\bigg|_{\eps=0}=\hat{\tau}_{ij}^a \hat{\tau}_{ij}^b l_{\langle ij\rangle}
\end{align}
after employing equation~\ref{eq:symmProp}. It's important to note that all of the columns of our constraint Jacobian are linearly dependent which prevents the QR decomposition algorithm from taking advantage of the speed increase gained by assuming that the constraints between different clusters are orthogonal.

As the strain tensor is required to be a $d\times d$ symmetric matrix, there are $d(d+1)/2$ strain degrees of freedom.
However, there is one final constraint to consider for shear stabilizing a packing of polymers: constraining the box volume. For a packing where the box volume is not constrained, the stress will always lead to the decompression of the box causing an unjamming event. It is therefore required that the applied strains keep the box volume constant. For infinitesimal strains, this constraint is simply $\Tr(\eps)=0$~\cite{torquato_breakdown_2003, donev_jamming_2004, donev_linear_2004}. The derivative of this constraint with respect to positional degrees of freedom is zero, but for the strain degrees of freedom, this means that
\begin{align}
\frac{\partial}{\partial \tensor{\eps}{^a_b}}\left(\sum_{\alpha}\tensor{\eps}{^\alpha_\alpha}\right)=\delta_{ab}.
\end{align}
To avoid confusion, the constraint Jacobian is referred to as $\Omega$ if it only involves positional degrees of freedom and $\Gamma$ if it involves both positional and strain degrees of freedom as well as the volume preserving constraint. Likewise, the othornormalized versions are called $\Omega_o$ and $\Gamma_o$ respectively. Just as in equation~\ref{constrainedForce}, given an arbitrary force vector that involves the derivatives of positions and strain, $\vec{F},$ the constrained force is given by
\begin{align}
\vec{F}_{\textrm{cons}}=\left(\mathbf{1}-\Gamma_o\Gamma_o^T\right)\vec{F}.
\end{align}
\subsection{Performing Minimization}
The constraints described above can be applied to any minimization algorithm that involves forces, such as gradient descent and FIRE~\cite{bitzek_structural_2006} as well as molecular dynamics simulations. While this algorithm can be realized with any potential, I choose to direct my attention to a soft sphere potential of the form,
\begin{align}
U(x,\eps)= \frac{1}{2w}\sum_i\sum_{j\in \partial i}\left(\xi_{ij}\right)^w
\end{align}
where $w$ is some power and $\xi_{ij}$ is the normalized overlap, 
\begin{align}
\xi_{ij}=1-\frac{\rho_{ij}}{\sigma_{ij}}.
\end{align}
In this equation $\rho_{ij}=\sqrt{\sum_{\alpha}\left(\psi_{ij}^\alpha\right)^2},$ $\sigma_{ij}$ is the sum of the radii,
and $j\in \partial i$ are defined to be those particles $j$ which are in contact with particle $i.$ This is understood to be a one-sided potential meaning that particles which are not overlapping do not interact. As a result, the contact network of our polymers changes throughout the minimization.

The forces are unchanged,
\begin{align}
F_{k}^{\mu}=-\frac{\partial U}{\partial x_{k}^{\mu}}\bigg|_{\eps=0}=\sum_{i\in\partial k}\xi_{ik}^{w-1}\frac{\hat{\tau}_{ik}^{\mu}}{\sigma_{ik}}
\end{align}
and the stresses are
\begin{align}
\sigma_{ab}=\frac{\partial U}{\partial \tensor{\eps}{^a_b}}\bigg|_{\eps = 0}=-\frac{1}{2}\sum_{i}\sum_{j\in \partial i}\frac{\rho_{ij}}{\sigma_{ij}}\xi_{ij}^{w-1}\hat{\tau}_{ij}^{a}\hat{\tau}_{ij}^{b}.
\end{align}

Now all that is left to do is to apply constraints to these forces and perform the minimization. After each minimization step, the positions and strains are updated. To simplify the scheme, the particle perturbations are applied first then the affine strain perturbs the particles further. The strain to our lattice vectors is also applied such that
\begin{align}
\Lambda_{\textrm{new}}=\left(\mathbf{1}+\eps\right)\Lambda
.\end{align} Finally, the strain is reset to zero. However, applying a finite strain to the lattice vectors will only preserve the volume to first order. To correct this, the lattice vectors are uniformly rescaled after each step so that they have a determinant of one. The polymer constraints also become violated to higher order so the same scheme that appears in section~\ref{sec:hoc} must be applied with our updated lattice vectors. To demonstrate the success of this methodology, two and three dimensional minimized system with $50$ rigid clusters each containing three particles is shown in Figure~\ref{fig:minimized}.

\subsection{Crossing Links, Rattling clusters, and Danglers}
\begin{figure}[]
\includegraphics[width=0.475\textwidth]{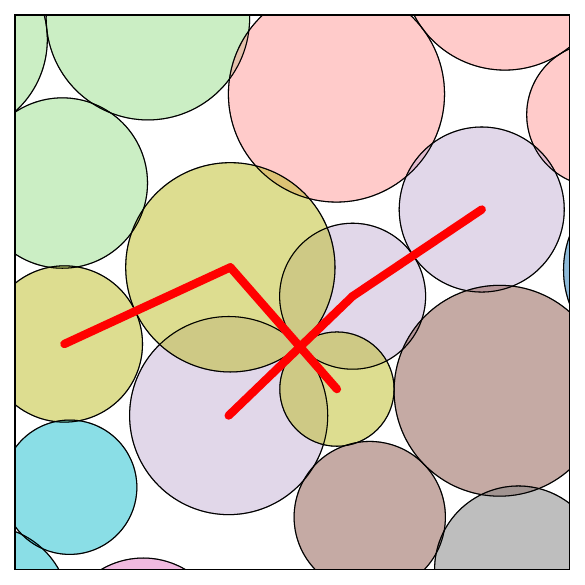}
\caption{Two clusters and their links are shown in a packing of short chains. Two of the red links were crossing before minimization and these two links remain crossing after the minimization is finished. These two chains are in force balance despite being in a high energy configuration. This only occurs in two dimensions and can be mitigated by ensuring that there are no crossing links before beginning the quench.}\label{fig:crossingLink}
\end{figure}

\begin{figure}[]
\includegraphics[width=0.475\textwidth]{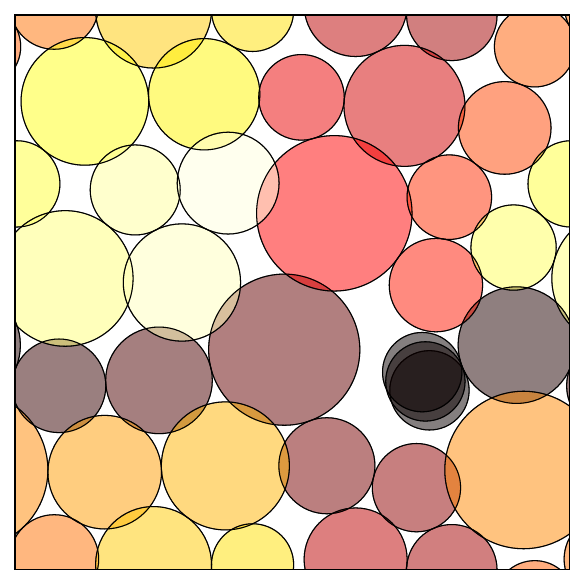}
\caption{A chain of 30 particles in force balance. This chain is an example of a rattling cluster because one end of the chain (black) is constrained by its link, but still has a degree of freedom to move (as shown). This free motion is a floppy mode and this means that the packing is not collectively jammed. However, simply removing the end of the chain will remove this floppy mode from the system, leaving a subsystem that is collectively jammed.}\label{fig:rattlers}
\end{figure}

Careful initialization of two dimensional packings is important to avoid crossing links. If two dimers have links that cross, as seen in Figure~\ref{fig:crossingLink}, this forms a stable configuration that minimization will not affect. This behavior can occur between different clusters or even in a single cluster of adequate length. To avoid this behavior, one can initialize the system such that link crossings are forbidden prior to minimization. However, if large overlaps are present in a configuration prior to minimization, link crossings may still occur. This behavior becomes more likely for larger timesteps in the beginning of the quench.
In a system of monomers at densities close to jamming, particles which are not locally constrained, termed rattlers, may introduce zero energy eigenmodes, or floppy modes, to the system. Rattlers are particles that are able to move independently of the other particles without affecting the system's energy. Polymer chain systems analogously can have rattling clusters. A rattling cluster is a cluster in which a subset of the cluster can move independently of the other clusters without affecting the system's energy. A particular type of rattling cluster that may appear is a dangler. A dangler is a single particle that does not interact with the other particles except by it's link (see Figure~\ref{fig:rattlers}).
\section{Finding Critically Jammed Systems}\label{sec:critJam}
A critically jammed packing is a packing that is rigid and has zero energy. To create such a polymer packing, start with randomly distributed cluster positions and bond angles (avoiding link crossings in 2D) at a density which is much higher than the expected jamming density. These clusters are allowed to interact via a harmonic contact potential and the energy is minimized with the FIRE algorithm. The size of each cluster and the corresponding constraints are then uniformly decreased while maintaining their average positions. After minimizing this lower-density packing, the process is repeated until the energy reaches a low threshold. The amount by which the packing fraction is decreased at each step is decided by taking advantage of the scaling relationship between energy and distance to the jamming density for soft sphere systems, $\phi-\phi_j,$ given in reference~\cite{charbonneau_jamming_2015}.

Because of the aforementioned rattling clusters, packings prepared in this manner will typically have some form of floppy mode present. This effect will be discussed in section~\ref{sec:strictJamming}.
\section{Defining the constrained Hessian and Rigidity Matrices}
\subsection{The Constrained Hessian}
For a packing of polymers, one may wish to find the normal modes, with or without strain degrees of freedom. In order to do that, one must first find the second derivatives of the energy function with respect to these degrees of freedom. These are as follows,
\begin{align}
&H_{xx}=\left.\frac{\partial^2 U}{\partial r_{j}^{\nu}\partial r_{k}^{\mu}}\right|_{\eps = 0}=\\
&\delta_{kj}\sum_{i\in \partial k}\left[\left(w-1\right)\frac{\xi_{ik}^{w-2}}{\sigma_{ik}^2}\hat{\tau}_{ik}^{\mu}\hat{\tau}_{ik}^{\nu}+\frac{\xi_{ik}^{w-1}}{\rho_{ik}\sigma_{ik}}(\hat{\tau}_{ik}^{\mu}\hat{\tau}_{ik}^{\nu}-\delta^{\mu\nu})\right]\nonumber\\
&-\delta_{\langle kj \rangle}\left[\left(w-1\right)\frac{\xi_{kj}^{w-2}}{\sigma_{kj}^2}\hat{\tau}_{kj}^{\mu}\hat{\tau}_{kj}^{\nu}+\frac{\xi_{kj}^{w-1}}{\rho_{kj}\sigma_{kj}}(\hat{\tau}_{kj}^{\mu}\hat{\tau}_{kj}^{\nu}-\delta^{\mu\nu})\right],\nonumber
\end{align}
\begin{align}
&H_{\eps x}=\left.\frac{\partial^2 U}{\partial \tensor{\eps}{^a_b}\partial r_{k}^{\mu}}\right|_{\eps=0}=\\
&\sum_{i}\frac{\delta_{\langle ik\rangle}}{\sigma_{ik}}\biggl\{\left[\left(w-1\right)\xi_{ik}^{w-2}-\left(w-2\right)\xi_{ik}^{w-1}\right]\hat{\tau}_{ik}^{\mu}\hat{\tau}_{ik}^{\kappa}\hat{\tau}_{ik}^{\beta}\biggr.\nonumber\\
&\biggl.-\xi_{ik}^{w-1}\left(\delta^{\beta\mu}\hat{\tau}_{ik}^{\kappa}+\delta^{\mu\kappa}\hat{\tau}_{ik}^{\beta}\right)\biggr\}\left(\frac{\partial \tensor{\eps}{^\kappa_\beta}}{\partial \tensor{\eps}{^a_b}}\right),\nonumber
\end{align}
and
\begin{align}
&H_{\eps\eps}=\left.\frac{\partial^2 U}{\partial \tensor{\eps}{^g_h} \partial \tensor{\eps}{^a_b}}\right|_{\eps=0}=\\
&\frac{1}{2}\sum_{i}\sum_{j\in\partial i}\left(1-\xi_{ij}\right)\biggl\{\left[\left(w-1\right)\xi_{ij}^{w-2}-\left(w-2\right)\xi_{ij}^{w-1}\right]\hat{\tau}_{ij}^{\alpha}\hat{\tau}_{ij}^{\mu}\biggr.\nonumber\\
&\biggl.-\xi_{ij}^{w-1}\delta^{\alpha\mu}\biggr\}\hat{\tau}_{ij}^{\beta}\hat{\tau}_{ij}^{\nu}\left(\frac{\partial\tensor{\eps}{^\mu_\nu}}{\partial\tensor{\eps}{^g_h}}\right)\left(\frac{\partial\tensor{\eps}{^\alpha_\beta}}{\partial\tensor{\eps}{^a_b}}\right).\nonumber
\end{align}
In these equations, the term $\delta_{\langle k m\rangle}$ is equal to one if particles $k$ and $m$ are in contact and zero otherwise.
These terms can be combined to find the extended hessian, which is the second derivative of the energy function in terms of both positional and strain degrees of freedom:
\begin{align}
H_0=\begin{pmatrix}
H_{xx} & H_{x\eps} \\ H_{x\eps}^T & H_{\eps\eps}
\end{pmatrix}.
\end{align}
The extended hessian can be used to find the energy of a perturbation that is done to the positions of individual particles and the strains. However, with polymer packings, we do not have access to all of these degrees of freedom. If there are $N$ particles in $d$ dimensions with $N_l$ nondegenerate links, the extended hessian will have $Nd+d(d+1)/2$ rows and columns whereas there are actually $N_{\textrm{dof}}=Nd+d(d+1)/2-N_l-1$ degrees of freedom (where the constraints due to the links and the volume-preserving strain have been subtracted). In order to calculate the energy of a perturbation and the normal modes of the polymer packing, one needs to translate the perturbations of the particles and affine strains to some basis of the true degrees of freedom. This can be achieved by performing a change of basis from the original basis to a basis of the true degrees of freedom and the constraints.

Let $\vec{y}$ be a vector of length $Nd+d(d+1)/2$ that contains the position and strain variables, let $\vec{y}_t$ be a vector of length $N_{\textrm{dof}}$ that contains the true degrees of freedom, and let $\vec{\lambda}$ be a vector of length $N_l+1$ corresponding to the constraint degrees of freedom. We need a square matrix, $Q,$ that decomposes $\vec{y}$ into $\vec{y}_t$ and $\vec{\lambda}$ such that
\begin{align}
\begin{pmatrix}
\vec{\lambda} \\ \vec{y}_t
\end{pmatrix}=Q^T\vec{y}.
\end{align}
Without loss of generality, one can define a matrix,
\begin{align}
\begin{pmatrix}
\Gamma & \mathrm{Null}(\Gamma^T)
\end{pmatrix}.
\end{align}
This gives a non-singular matrix where the first $N_l+1$ columns correspond to our constraints. This matrix can then be subjected to QR decomposition to give a matrix $Q.$

With this new matrix, $Q,$ one can define a rectangular change of basis matrix as
\begin{align}
B=\begin{pmatrix}
\mathbf{0}_{N_l+1} & \mathbf{1}_{N_{\textrm{dof}}}
\end{pmatrix}Q^T
\end{align}
such that
\begin{align}
\vec{y}_t=B\vec{y}.
\end{align}
This basis is also useful for removing the components of a vector $\vec{y}$ that violate our constraints:
\begin{align}
\vec{y}'=B^TB\vec{y}.\label{eq:removeComp}
\end{align}
With this matrix, $B,$ the constrained extended hessian becomes
\begin{align}
H_\textrm{E}=BH_0B^T.
\end{align}
Given some perturbation, $\vec{y}_t$ of our $N_{\textrm{dof}}$ degrees of freedom, the change in energy can be computed as
\begin{align}
\Delta E=\frac{1}{2}\vec{y}_t^TH_\textrm{E}\vec{y}_t.
\end{align}
The extended hessian can also be diagonalized to find the normal modes. The only problem is that the normal modes are in terms of a rather confusing basis, but this can be rectified by taking the matrix of eigenvectors and multiplying them by $B^T$ giving a set of $N_\textrm{dof}$ eigenmodes in the familiar basis of positions of particles and strains. This entire procedure can also be adapted to use the matrix $\Omega$ instead of $\Gamma.$ This will create an extended hessian that deals only with positional degrees of freedom.

\begin{figure}[]
\includegraphics[width=0.475\textwidth]{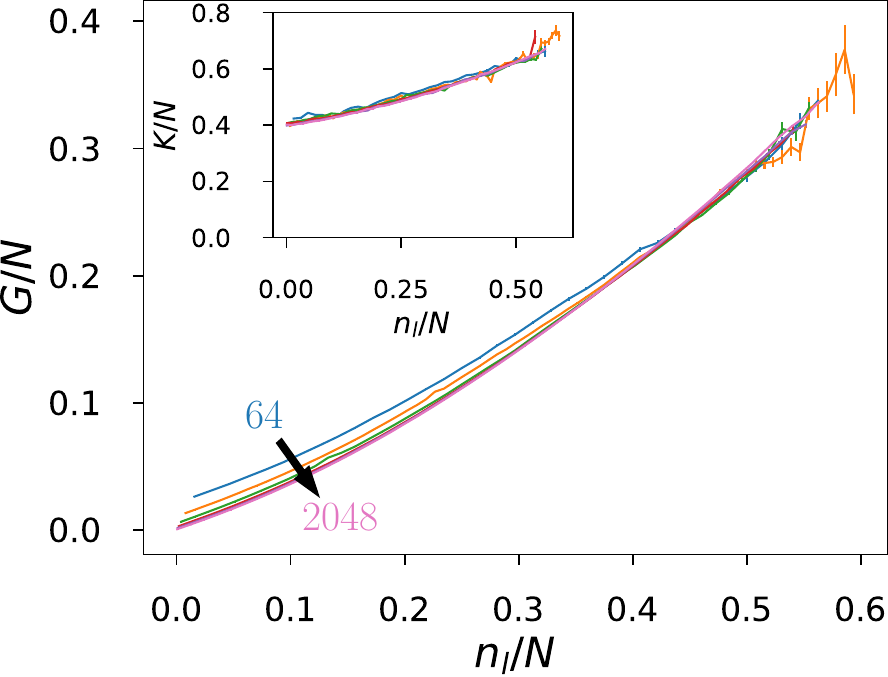}
\caption{The shear and bulk (inset) moduli of three dimensional packings of clustered particles. Starting with shear stabilized, critically jammed systems of monomers, links are randomly added to the contact network and bond angles are frozen to simulate cementing events. The number of links at any given point is $n_l$ and this does not include those links which are added to fix the bond angles. There are about $25$ packings for each $N$ (from $2^6$ to $2^11$) in which $50$ independent percolation experiments were performed. The data for each $N$ and $n_l$ was subject to a weighted average where the Reuss and Voigt averages were used as a minimum and maximum respectively. The error bars show the weighted standard error.}\label{fig:shearBulkFig}
\end{figure}

\subsection{The Constrained Rigidity Matrix}
If one wanted to examine the underlying unnormalized, unstressed spring network of a packing and find the states of self stress, they could define the extended rigidity matrix. The rigidity matrix relates perturbations to bond stresses, so to derive it, consider the effect that perturbing or straining the packing has on the bond stresses. For particle perturbations, the rigidity matrix has the form
\begin{align}
R_{x,\langle ij \rangle \left(k\gamma\right)}=\left(\delta_{jk}-\delta_{ik}\right)\hat{\tau}_{ij}^{\gamma}.
\end{align}
For strains, the rigidity matrix has the form
\begin{align}
R_{\eps,\langle ij \rangle \left(k\gamma\right)}=\hat{\tau}_{ij}^{\alpha}\hat{\tau}_{ij}^{\beta}\sigma_{ij}.
\end{align}
These two terms can be combined to get the rigidity matrix in terms of strains and positions,
\begin{align}
R_{0, \langle ij \rangle}=\begin{pmatrix}
R_{x,\langle ij \rangle} & R_{\eps,\langle ij \rangle}
\end{pmatrix}
\end{align}
such that for a vector $\vec{y}=\begin{pmatrix}
\vec{x} \\ \vec{\eps}
\end{pmatrix},$
\begin{align}
R_{0, \langle ij \rangle}\vec{y}=R_{x,\langle ij \rangle}\vec{x} + R_{\eps,\langle ij \rangle}\vec{\eps}.
\end{align}
As before, one can perform a change of basis on this rigidity matrix to find the constrained extended rigidity matrix,
\begin{align}
R_{\textrm{E}}=R_0B^T.
\end{align}
Note that the constrained extended rigidity matrix is only defined for the bonds in the system, not for the links. If the links were included, then $R_{\textrm{E}}$ would return zero stress on those bonds regardless of the choice for $\vec{y}_t.$

Similarly, the states of self stress for the network are the left singular vectors that have a zero singular value. The constrained extended $\mathcal{N}$-matrix, which can be used to find the states of self stress, can be computed as $\mathcal{N}_{\textrm{E}}=R_{\textrm{E}}R_{\textrm E}^T.$ Likewise, the constrained extended dynamical matrix as $\mathcal{D}_{\textrm{E}}=R_{\textrm{E}}^TR_{\textrm E}$ for the underlying unstressed spring network.
\section{Testing for strict jamming}\label{sec:strictJamming}
If one were to make a hard sphere polymer packing, such as those found by following the procedure described in section~\ref{sec:critJam}, one might want to know whether or not this packing remains stable against all possible combinations of strains and perturbations. One way to do this is to employ a linear programming algorithm based on the one found in reference~\cite{donev_linear_2004} with the constrained extended rigidity matrix in place of the adjacency matrix. The linear program is:
\begin{align}
&\textrm{min~}\vec{b}^T\vec{y}_t\\
\textrm{such that~}&R_{\textrm{E}}\vec{y}_t\leq \vec{0}\\
\textrm{where~}&\left|\vec{y}_t\right|\leq y_{\textrm{max}}.
\end{align}
In this program, we are looking for the vector $\vec{y}_t$ which is subjected to some random load vector $\vec{b}$ that is bounded such that the length is less than some finite value $y_{\textrm{max}}.$ If this algorithm returns a nonzero vector, $\vec{y}_t,$ then $\vec{y}_t$ describes an unjamming motion. Because of the presence of rattling clusters, this may be the case. To determine which particles are contributing to the nonzero $\vec{y}_t,$ one can find the nonzero indices for $B^T\vec{y}_t.$ Those rattling clusters should be removed from the packing before the linear program is executed again. This process can be repeating until $\vec{y}_t=\vec{0}$ is found. One must also run the same linear program for $\textrm{min~}-\vec{b}^T\vec{y}_t$ to ensure that the polymer subpacking is strictly jammed~\cite{donev_linear_2004} since the vector $\vec{b}$ is defined arbitrarily. As in the previous sections, the same process can be adapted to $R_{x}$ to test for collective jamming.

\section{Computing the Compliance Matrix}
Now that jamming and normal modes for the polymer systems have been discussed, the discussion can conclude by computing the elastic moduli. To find the elastic moduli, the compliance matrix, $S,$ needs to be defined. This matrix relates the stress to the strain,
\begin{align}
\vec{\eps}=S\vec{\sigma}.
\end{align}
Before this is derived, consider Hooke's law for the unconstrained extended hessian,
\begin{align}
H_0\begin{pmatrix}
\Delta\vec{x} \\ \vec{\eps}
\end{pmatrix}=\begin{pmatrix}
-\vec{F} \\ \vec{\sigma}
\end{pmatrix}.
\end{align}
Applying an arbitrary strain, $\vec{\eps},$ and perturbation, $\Delta\vec{x},$ puts a stress, $\vec{\sigma},$ and interparticle forces, $\vec{F},$ on our packing. Not every combination of $\Delta \vec{x}$ and $\vec{\eps}$ is allowed, so we need to project out the part of our vector that violates the constraints. From equation~\ref{eq:removeComp} one can achieve this with $B^TB.$ However, this is not quite correct. When finding the elastic moduli, deformations which may affect the volume of the packing are allowed. As such, $B$ is rederived with the volume-conserving constraint excluded from $\Gamma.$ This new rectangular change of basis is referred to as $B_{-1}.$ A new constrained hessian can be defined as
\begin{align}
H_c=B_{-1}^TB_{-1}H_0B_{-1}^TB_{-1}.\label{eq:block}
\end{align}

To find the stress-strain relationship, it is not enough to apply an affine strain. Simply applying an affine strain will cause the packing to lose force balance. When the stress-strain relationship is probed in granular packings, minimization steps are taken between strain steps. What one must do is apply an arbitrary affine strain and a corresponding nonaffine perturbation, $\Delta \vec{x}_{\textrm{na}},$ such that force balance is kept. For an unconstrained hessian, Hooke's law can be applied to achieve an equation such as the following:
\begin{align}
H_0\begin{pmatrix}
\Delta \vec{x}_{\textrm{na}} \\ \vec{\eps}
\end{pmatrix}=\begin{pmatrix}
\vec{0} \\ \vec{\sigma}
\end{pmatrix}.
\end{align}
However, for the constrained hessian, this relationship is false. To understand why, imagine applying a particular perturbation and strain that strictly violate our constraints; this would result in zero strain. This is the exact opposite of what one would expect. It should be impossible to apply such a perturbation and strain, therefore one would expect the result of such a test to return an infinite stress. This is remedied by taking the Moore-Penrose pseudoinverse~\cite{ben-israel_existence_2003} of $H_c.$ This works because the pseudoinverse preserves all of the zero eigenvalues. We can then conclude that
\begin{align}
\left(H_c\right)^{-1}\begin{pmatrix}
\vec{0} \\ \vec{\sigma}
\end{pmatrix}=\begin{pmatrix}
\Delta \vec{x}_{\textrm{na}} \\ \vec{\eps}
\end{pmatrix}.
\end{align}
This is much easier to understand as well because while certain strains may not be possible, any stress is allowed. The result will never violate our constraints, but may lead to zero strain.
If $\left(H_c\right)^{-1}$ is partitioned,
\begin{align}
\vec{\eps}=S\vec{\sigma}
\end{align}
where
\begin{align}
S = \begin{pmatrix}
\mathbf{0}_{s\times N_p} & \mathbf{1}_{s\times s}
\end{pmatrix}\left(H_c\right)^{-1}\begin{pmatrix}
\mathbf{0}_{N_p\times s} \\ \mathbf{1}_{s\times s}
\end{pmatrix}
\end{align}
is the compliance matrix where $N_p$ is the number of particle degrees of freedom and $s$ is the number of strain degrees of freedom.

For three dimensional polymer chain systems, the shear and bulk moduli can be found from the Voigt, Reuss, and Hill averages under the assumption that the configuration is nearly isotropic~\cite{hill_elastic_1952, watt_elastic_1976}. From these averages, the poisson ratio and anisotropy can also be calculated. To demonstrate this procedure, I consider three dimensional shear stabilized systems of monodisperse monomers at a single state of self stress and randomly replace some of their contacts with links. At a certain point sufficiently many links are added to prevent certain stresses from causing strains. These impossible stresses show up as zero modes in the compliance matrix and indicate a direction in which the shear modulus is infinite. Adding additional links will also eventually cause the compliance matrix to become zero, indicating a non-deformable packing of polymer chains. The bulk and shear moduli for these systems as a function of the fraction of added links, $n_l/N,$ (not including those added to preserve bond angles) are shown in Figure~\ref{fig:shearBulkFig}.

\section{Conclusions}
In this methods paper I have discussed how to generate packings of arbitrarily defined polymer chains. I described how to simulate the annealing of these packings and how they can be shear stabilized in the process. I gave examples of undesirable behaviors and how to prevent them as well as the definitions of rattling clusters and danglers. I then explained how to find the normal modes, classification in the jamming hierarchy, and elastic moduli. This work lays the foundations for a more thorough exploration of the mechanical properties of packings of polymer chains and molecules as well as a clear method for furthering our understanding of many important topics such as the polymer glass transition, clumping, and cementing events. Additionally, this basic framework can be altered to constrain configurations in other ways such as generating area and/or perimeter preserving polytopes, molecules that lack rotational degrees of freedom, or even rigid bistable structures.

\section{Acknowledgments}
I thank Eric Corwin, James Sartor, and Heinrich Jaeger for helpful discussions and feedback. This work was supported by National Science Foundation (NSF) Career Award DMR-1255370 and the Simons Foundation No. 454939.

\bibliography{molecularRigidity}

\end{document}